\documentclass[sigconf,pbalance]{acmart}

\pdfoutput=1
\usepackage[utf8]{inputenc}

\setcopyright{none}
\renewcommand\footnotetextcopyrightpermission[1]{} 
\usepackage[T1]{fontenc}
\usepackage{url}
\usepackage{flushend}
\usepackage{xspace}
\usepackage[skip=5pt]{caption}
\usepackage[inline]{enumitem}
\usepackage{ifthen}
\usepackage{balance}
\usepackage{pifont}

\usepackage[frozencache]{minted}
\setminted{breaklines,breakanywhere}
\setmintedinline{breaklines,breakafter=.}

\usepackage{microtype}
\usepackage{multirow}
\usepackage[]{numprint}
\newcommand*\np[2][z]{
\ifx z#1%
$\numprint{#2}$%
\else%
$\numprint[#1]{#2}$%
\fi\xspace%
}

\usepackage{fp}
\usepackage{xfp}

\usepackage{graphicx}
\usepackage{subcaption}
\usepackage{pifont}

\usepackage{booktabs}
\usepackage{multirow}
\usepackage{threeparttable}
\usepackage{xcolor}
\usepackage{array,makecell}
\usepackage{diagbox}
\usepackage{tabularx}

\usepackage{mathtools}
\usepackage{amsmath}

\usepackage{tcolorbox}
\usepackage{pgfplots}
\usepackage{pgfplotstable}

\usepackage{enumitem}
\usepackage[framemethod=TikZ]{mdframed}
\usepackage[scaled]{beramono} 

\usepgfplotslibrary{statistics}
\usetikzlibrary{calc,patterns,fpu} 
\pgfplotsset{compat=1.14} 

\pgfplotsset{
    /pgf/declare function={
        Floor(\x) = round(\x-0.49);
    },
    show sum on top/.style={
        /pgfplots/scatter/@post marker code/.append code={%
            \path let \p1=($(normalized axis cs:%
                        \pgfkeysvalueof{/data point/x},%
                        \pgfkeysvalueof{/data point/y})%
                        -(normalized axis cs:\pgfkeysvalueof{/data point/x},0)$)
            in node[
                at={(normalized axis cs:%
                        \pgfkeysvalueof{/data point/x},%
                        \pgfkeysvalueof{/data point/y})%
                },
                anchor={-90*sign(\y1)},yshift={sign(\y1)*2pt}
            ]
            {\pgfmathprintnumber[fixed, precision=1]{\pgfkeysvalueof{/data point/y}}};
        },
    }
}

\definecolor{blau_1a}{RGB}{93,133,195}
\definecolor{blau_2a}{RGB}{0,156,218}
\definecolor{gruen_3a}{RGB}{80,182,149}
\definecolor{gruen_4a}{RGB}{175,204,80}
\definecolor{gruen_5a}{RGB}{221,223,72}
\definecolor{orange_6a}{RGB}{255,224,92}
\definecolor{orange_7a}{RGB}{248,186,60}
\definecolor{rot_8a}{RGB}{238,122,52}
\definecolor{rot_9a}{RGB}{233,80,62}
\definecolor{lila_10a}{RGB}{201,48,142}
\definecolor{lila_11a}{RGB}{128,69,151}

\definecolor{blau_1b}{RGB}{0,90,169}
\definecolor{blau_2b}{RGB}{0,131,204}
\definecolor{gruen_3b}{RGB}{0,157,129}
\definecolor{gruen_4b}{RGB}{153,192,0}
\definecolor{gruen_5b}{RGB}{201,212,0}
\definecolor{orange_6b}{RGB}{253,202,0}
\definecolor{orange_7b}{RGB}{245,163,0}
\definecolor{rot_8b}{RGB}{236,101,0}
\definecolor{rot_9b}{RGB}{230,0,26}
\definecolor{lila_10b}{RGB}{166,0,132}
\definecolor{lila_11b}{RGB}{114,16,133}

\newboolean{showcomments}
\setboolean{showcomments}{true}

\newcommand{\ShowAbsoluteNumber}[1]{%
\ifnum #1<10%
{\hspace*{0pt}#1}%
\else%
\ifnum #1<100%
{\hspace*{0pt}#1}%
\else%
\ifnum #1<1000%
{\hspace*{0pt}#1}%
\else%
{\numprint{#1}}%
\fi%
\fi%
\fi%
}

\newcommand{\ShowPercentage}[2]{%
\FPeval\percentage{round(#1/#2*100,0)}%
\FPeval\percentageOneDecimal{round(#1/#2*100,1)}%
\ifnum \percentage=0%
{\np[\%]{\FPprint{percentageOneDecimal}}}%
\else%
\ifnum \percentage<10%
{\np[\%]{\FPprint{percentageOneDecimal}}}%
\else%
{\np[\%]{\FPprint{percentageOneDecimal}}}%
\fi%
\fi%
\xspace
}
\newlength\BARSIZE  
\newcommand{\inlinechart}[2]{%
\FPeval{\BLACKBARSIZE}{#1/#2}\textcolor{black!80}{\rule{\BLACKBARSIZE\BARSIZE}{1.6ex}}%
\FPeval{\BLACKBARSIZE}{1 - (#1/#2)}\textcolor{black!10}{\rule{\BLACKBARSIZE\BARSIZE}{1.6ex}}%
}

\newcommand*\percent[3][v]{%
\ifx q#1%
    \np{#2}/\np{#3}(\ShowPercentage{#2}{#3})\else%
\ifx s#1%
    \ShowPercentage{#2}{#3}\else%
\ifx p#1%
    \np{#2}(\ShowPercentage{#2}{#3})\else%
\ifx c#1%
    \inlinechart{#2}{#3}%
\else%
    \np{#2}%
    \ifx r#1%
        /\np{#3}%
    \fi%
    \hspace*{0.5ex}(\ShowPercentage{#2}{#3}) %
    \inlinechart{#2}{#3}%
    \xspace
\fi\fi\fi\fi%
}
\def\tool{\textit{{EnergiBridge}}\xspace}

\makeatletter
\@ifclasswith{acmart}{anonymous}{%
\def\toolURL{}%
}{%
\def\toolURL{https://github.com/tdurieux/EnergiBridge}%
}
\@ifclasswith{acmart}{anonymous}{%
\def\replicationURL{}%
}{%
\def\replicationURL{https://github.com/tdurieux/EnergiBridge-experiments}%
}
\makeatother

\definecolor{mygray}{RGB}{240,240,240}

\ifthenelse{\boolean{showcomments}}
 { }
        { }

\definecolor{eminence}{RGB}{108,48,130}
\definecolor{weborange}{RGB}{255,165,0}
\definecolor{frenchplum}{RGB}{129,20,82}
\definecolor{darkgreen}{RGB}{10, 92, 10}

\mdfdefinestyle{mpdframe}{
    nobreak                     =true,
    backgroundcolor             =black!3,
    frametitlebackgroundcolor   =black!15,
    frametitlerule              =false,
    roundcorner                 =5pt,
    innermargin                 =0.3cm,
    innerleftmargin             =0.3cm,
    innerrightmargin            =0.3cm,
    innertopmargin              =0.3cm,
    innerbottommargin           =0.3cm,
    shadowsize                  =1pt
}

\usepackage{bbding}
\newcommand\no{}
\newcommand\YES{\scriptsize\Checkmark}

\usepackage{xspace}
\newcommand{\ie}{\emph{i.e.,}\xspace}
\newcommand{\eg}{\emph{e.g.,}\xspace}

\title{\tool: Empowering Software Sustainability through Cross-Platform Energy Measurement}

\author{June Sallou}
\affiliation{%
    \institution{TU Delft}
    \country{The Netherlands}
}
\email{J.Sallou@tudelft.nl}
\orcid{0000-0003-2230-9351}

\author{Luís Cruz}
\affiliation{%
    \institution{TU Delft}
    \country{The Netherlands}
}
\email{L.Cruz@tudelft.nl}
\orcid{0000-0002-1615-355X}

\author{Thomas Durieux}
\affiliation{%
    \institution{TU Delft}
    \country{The Netherlands}
}
\email{thomas@durieux.me}
\orcid{0000-0002-1996-6134}


\begin{document}

\begin{abstract}
In the continually evolving realm of software engineering, the need to address software energy consumption has gained increasing prominence. However, the absence of a platform-independent tool that facilitates straightforward energy measurements remains a notable gap. This paper presents \tool, a cross-platform measurement utility that provides support for Linux, Windows, and MacOS, as well as Intel, AMD, and Apple ARM CPU architectures.
In essence, \tool serves as a bridge between energy-conscious software engineering and the diverse software environments in which it operates. It encourages a broader community to make informed decisions, minimize energy consumption, and reduce the environmental impact of software systems. 

By simplifying software energy measurements, \tool offers a valuable resource to make green software development more lightweight, education more inclusive, and research more reproducible. Through the evaluation, we highlight \tool's ability to gather energy data across diverse platforms and hardware configurations.

\tool is publicly available on GitHub: \url{\toolURL}, and a demonstration video can be viewed at: \url{https://youtu.be/-gPJurKFraE}.
\end{abstract}

\maketitle

\section{Introduction}

Software engineering is a dynamic and ever-evolving field, focused on creating, maintaining, and evolving software systems to meet the ever-growing demands of modern society. A fundamental aspect of this discipline is to gauge the evolution of software and ensure its ongoing reliability and efficiency. As software projects grow in complexity and scale, it becomes imperative to employ rigorous metrics to assess their quality, performance, and sustainability.

Among the metrics that have gained prominence in the last decade are code coverage and linter scores.
Moreover, as the world increasingly turns its attention to energy conservation and sustainability, the software engineering landscape is adapting to consider these crucial aspects as well~\cite{Moises2018, Jagroep2017}. 
Understanding and measuring the energy consumption of software has become a pressing concern, with implications for environmental impact, cost efficiency, and the overall sustainability of software systems \cite{Guldner2021Oct}.

One significant challenge in this endeavor is the diversity of the environments in which software operates. Software is executed on a wide range of operating systems and hardware configurations, making it difficult to employ uniform measurement techniques. 
Often, existing approaches are limited to specific processor brands or rely on specialized and complex energy consumption APIs that frequently lack comprehensive documentation, as seen with RAPL~\cite{linux-power-interfaces}. 
This complexity makes it difficult to be adopted by practitioners, who generally lack the knowledge for such measurements~\cite{Pinto2017, Pang2015Jul}. 

To address these challenges, we introduce \tool, a cross-platform energy measurement tool. 
\tool is designed to support multiple operating systems, including Linux, Windows, and Mac OS, as well as a wide spectrum of CPU architectures: Intel, AMD, and Apple ARM. 
Our primary goal is to simplify the process of measuring energy consumption across diverse platforms, providing software engineers with a unified toolset, at a relevant abstraction level, for assessing the energy consumption of their software.
The goal is also to provide a tool that can be easily used in education to promote the energy aspect of software.

The ultimate objective of \tool is to democratize the consideration of energy consumption in software development. We envision \tool as a user-friendly solution that empowers a broader community of software practitioners to incorporate energy efficiency as a fundamental criterion in their software design and development processes. 
By bridging the gap between energy-conscious software engineering and the diverse technology landscape, we aspire to contribute to a more sustainable and environmentally responsible software ecosystem.

The contributions of this paper are the following:
\begin{itemize}
    \item \tool: a tool to measure energy consumption in different ecosystems.
    \item A replication package that allows the replication of the energy measurement of Chrome execution. 
    \item Ready-to-use setup to measure the energy consumption of different applications.
\end{itemize}

\tool is publicly available on our repository: \url{\toolURL} and the replication package is available at \url{\replicationURL}.

\section{Related Works}\label{sec:background}

In contrast to physical power monitors, which are hardware-based external meters and very meticulous to set up (\eg requiring custom physical connections between the device or component under study and the power monitor, as well as time synchronization between software execution and monitoring), energy profilers are software-based energy measurement tools that are intended to be easier to set up~\cite{cruz2021tools}. However, they often operate only under very specific environments and are limited to one or two platforms or processor architectures. Hence, they limit their usage to a specific execution context.
Additionally, the tool output is not standardized and needs to be adapted for each analyzed platform and CPU.

\autoref{tab:enery_tool} presents the tools that we identify as the most commonly used for collecting energy consumption metrics, and their platform applicability.
This table illustrates the need for a cross-platform tool that can gather energy measurement data on different platforms and support the developers. 
This constant is reinforced by the fact that some of those tools are now deprecated by Intel.

Compared to the state of the art, \tool offers significantly broader support for operating systems and CPU platforms, making it compatible with a much larger population.

\begin{table}[!ht]
    \caption{Energy measurement tools and their compatibility with platforms.}
    \label{tab:enery_tool}
    \centering
    \setlength\tabcolsep{1.5pt}
    \begin{tabular}{@{}l|ccc|cc|cc@{}}
\toprule    
\multirow{2}{*}{Tool}& \multicolumn{3}{c|}{Windows} & \multicolumn{2}{c|}{Linux} & \multicolumn{2}{c}{Mac OS}\\\cline{2-8}
         & Intel & AMD & ARM& Intel & AMD &  Intel & ARM \\
 \midrule
\href{https://software.intel.com/content/www/us/en/develop/articles/intel-power-gadget.html}{Intel Power Gadget} & \YES & \no & \no & \no & \no  & \YES & \no \\
\href{https://software.intel.com/content/www/us/en/develop/articles/intel-power-gadget.html}{Intel PowerLog} & \YES & \no & \no & \no & \no  & \YES & \no \\
\href{https://www.seense.com/menubarstats/mxpg/}{Mx Power Gadget} & \no & \no & \no & \no & \no  & \no & \YES \\
\href{http://manpages.ubuntu.com/manpages/bionic/man8/powerstat.8.html}{Powerstat} & \no & \no & \no & \YES & \no  & \no & \no \\
\href{https://01.org/powertop}{PowerTOP} & \no & \no & \no & \YES & \YES  & \no & \no \\
\href{https://www.man7.org/linux/man-pages/man1/perf.1.html}{Perf} & \no & \no & \no & \YES & \YES  & \no & \no \\
\href{https://hpc.fau.de/research/tools/likwid/}{Likwid} & \no & \no & \no & \YES & \no  & \no & \no \\
\href{https://github.com/s4y/syspower/}{Syspower} & \no & \no & \no & \no & \no & \YES & \no \\
\href{https://learn.microsoft.com/en-us/windows-hardware/design/device-experiences/powercfg-command-line-options}{Powercfg} & \YES & \YES & \YES & \no & \no & \no & \no \\
\midrule
\href{\toolURL}{\textbf{\tool}} & \YES & \YES & \no & \YES & \YES & \YES & \YES \\
\bottomrule
\end{tabular}
\end{table}

\section{Purposes}

\tool presents a versatile range of use cases regarding its usage purposes, making it a valuable asset for various scenarios within the software engineering domain.

\textbf{1. Research and Practice:} \tool is designed to be easy to use, catering to the needs of both researchers and practitioners in the software engineering community. Researchers can use it to study the energy profiles of software applications, leading to a deeper understanding of performance bottlenecks and optimization opportunities. \tool enables reproducibility of such studies in different setups or environments without requiring any change on the tools being used or the format of the energy data being analyzed. Practitioners, on the other hand, can seamlessly integrate \tool into their development workflows to monitor and enhance the energy efficiency of their software products, ultimately reducing operational costs and environmental impact.

\textbf{2. Educational Initiatives:} In educational settings, students use a variety of platforms which makes it complex to provide a unique tool for everyone. Moreover, educators do not always have access to devices with all the same setups that students use. The alternative of enforcing a unique environment is also suboptimal, as it is less inclusive for students who are not able to get the same environment on their personal computers. 
\tool addresses this problem, and lets the education focus on the impact of software consumption instead of spending time discussing how to install different software or how to interpret the different outputs. 
It simplifies energy measurement and analysis, making it an excellent educational tool for instructors and students alike.

\textbf{3. Cross-Platform Energy Collection:} A use case for \tool is enabling the collection of software energy consumption across diverse platforms. 
This use case allows developers to analyze the behavior of their software on different operating systems and hardware configurations.
It is not necessarily possible to compare the measured value but it is possible to compare the general energy curve of the system as well as the evolution of the energy consumption of the software.

The versatility of \tool extends its utility across these diverse use cases and purposes, underscoring its significance as a valuable resource in the software engineering landscape.

\section{\tool}\label{sec:tool}

\tool is a user-friendly tool designed for seamless integration by developers.
The interface and output are uniform across platforms, therefore simplifying the integration and training process.
We also provide the executable of \tool for Windows, Linux, and MacOS directly through GitHub releases, making it accessible to new users and reducing the initial setup overhead.

To achieve this uniform interface, \tool is developed in Rust and relies on low-level system calls to collect the energy consumption and reduce the execution overhead that would emerge from merging existing tools together.
On Mac, it relies on \texttt{System Management Controller} (SMC) to read CPU temperature and the system power usage.
For Intel and AMD CPU, it accesses energy information from the CPU by reading the \texttt{Model-specific register} (MSR).
Intel and AMD report the energy in joules while Apple reports the average power in watts. 
For simplified and uniform usage, we convert all the information into energy using watts -- the SI unit for energy consumption per second.

\subsection{Usage}

The core functionality of \tool revolves around its command-line interface.
Developers can effortlessly invoke the tool by specifying a command to run as a parameter. 
During the execution of this command, \tool performs energy measurements at specific intervals. 
This approach allows for accurate and informative energy consumption analysis. \tool is used as follows:
\begin{minted}{text}
./energibridge[.exe] [OPTIONS] -- [COMMAND]
\end{minted}
Where the options can be:
\begin{itemize}[leftmargin=10pt]
\item \textbf{-i, --interval <INTERVAL>} Duration of the interval between two measurements in milliseconds [default: 100]
\item  \textbf{-m, --max-execution <MAX\_EXECUTION>} The maximum duration of the command execution in seconds. [default: 0]
\item \textbf{--summary} Display a summary of the energy consumption during the execution of the command.
\end{itemize}

In the example below, we demonstrate how to use \tool to measure the energy consumption of Chrome while running it for 10 seconds:
\begin{minted}{text}
./energibridge -m 10 -- google-chrome google.com
\end{minted}

\subsection{Output}

The output generated by \tool is designed with analysis in mind.
Instead of providing direct usability, we furnish a CSV file that contains measurements acquired at specified intervals.

Since each system is different, we collect varying information depending on the system, in particular regarding the granularity of the measurements.
An AMD CPU on Linux will be able to report the energy consumption of each CPU core while a ARM Mac will only be able to provide the system power consumption. 
\autoref{tab:cpu_support} presents the different CPU properties and \autoref{tab:gpu_support} presents the GPU properties that are collected on each system by \tool.

\begin{table}[!ht]
\caption{CPU Properties that are supported for each system.}\label{tab:cpu_support}
\setlength\tabcolsep{1.5pt}
\begin{tabular}{@{}l|ccc|ccc|cc@{}}
\toprule
\multirow{2}{*}{Property}& \multicolumn{3}{c|}{Windows} & \multicolumn{3}{c|}{Linux} & \multicolumn{2}{c}{Mac OS} \\\cline{2-9}
         & Intel & AMD & ARM& Intel & AMD & ARM&  Intel  & ARM \\
\midrule
CPU Usage      & \YES & \YES & \YES & \YES & \YES & \YES & \YES & \YES \\
Package Power  & \YES & \YES & \YES & \YES & \YES & \YES & \YES & \no \\
System Power   & \no & \no & \no & \no & \no & \no & \YES & \YES \\
Core Frequency  & \YES & \YES & \YES & \YES & \YES & \YES & \no & \no \\
Core Power     & \no & \no & \no & \no & \YES & \no & \no & \no \\
Memory Usage   & \YES & \YES & \YES & \YES & \YES & \YES & \YES & \YES \\
\bottomrule
\end{tabular}%
\end{table}

\begin{table}[!ht]
\caption{GPU Properties that are supported for each system.}\label{tab:gpu_support}
\begin{tabular}{@{}l|c|c|cc@{}}
\toprule
\multirow{2}{*}{Property}& \multicolumn{1}{c|}{Windows} & \multicolumn{1}{c|}{Linux} & \multicolumn{2}{c}{Mac OS} \\\cline{2-5}
         & Nvidia & Nvidia &  AMD/Intel  & ARM \\
\midrule
GPU Usage      & \YES & \YES & \no & \no  \\
GPU Frequency  & \YES & \YES & \no & \no \\
GPU Power  & \YES & \YES & \YES & \YES \\
\bottomrule
    \end{tabular}
\end{table}

\section{Evaluation}
In this section, we present the evaluation of \tool. To do so, we present a use case of \tool on a cross-platform system.
The main purpose of the evaluation is to illustrate the ability of \tool to capture the energy consumption of software on a cross-platform environment with a wide variety of hardware and software.

\subsection{Experimental use case}
The evaluation consists of running Chrome for a defined period and measuring its energy consumption and we will measure the difference with the same system running at IDLE (no other programs are running). We select Chrome as it is the most popular browser\footnote{\url{https://www.statista.com/statistics/268299/most-popular-internet-browsers/}} and multi-platform.
The Chrome workload consists of opening new Chrome windows, going to \textit{google.com}, and waiting for 10 seconds. 
The IDLE workload consists of measuring the energy consumption for 10 seconds while the system is doing nothing. 
\tool (\href{https://github.com/tdurieux/EnergiBridge/releases/tag/v0.0.4}{v0.0.4}, sha: \href{https://github.com/tdurieux/EnergiBridge/commit/dab662f77fc0fb92a4bcc91e20b6e5394e4a8410}{dab662f}) is used as follows:
\begin{minted}{text}
./energibridge[.exe] --max-execution 10 --output energy.csv --command-output Chrome.log --iterval 100 -- google-chrome google.com
\end{minted}

\autoref{tab:evaluation_harware} presents the hardware that is used for this evaluation. We aim to use different operating systems and hardware components.
During the execution of the evaluation, those systems had all their applications closed and no users were interacting with them.

\begin{table}[t]
    \centering
    \caption{Description of the configuration used during the evaluation.}
    \label{tab:evaluation_harware}
    \begin{tabular}{@{}lll@{}}
    \toprule
OS         &  CPU & GPU \\
\midrule
Windows   & AMD Ryzen™ 5 5600H & GeForce RTX 3060 Laptop \\
Linux     & AMD Ryzen™ 9 7900X & GeForce RTX 4090 \\
Mac ARM   & Apple M1 Max  & Apple M1 Max \\
Mac Intel & Intel Core i7-9750H & Intel Graphics 630 \\
\bottomrule
    \end{tabular}
\end{table}

We execute those two workloads (\ie Chrome and IDLE) 20 times with a random order on each system to collect statistically valid measurements. 
It is considered a good practice to run the measurements multiple times to ensure that no special situation happens during the collection (\eg background execution). 

Additionally, we provide the replication package of our evaluation \footnote{Replication package: \url{\replicationURL}}.
The replication package provides the experiment scripts, collected data, and associated analysis of the experimental use case. The same repository also includes an execution framework for measuring energy across various other application use cases, from browser execution to Large Language Model inference with LLAMA.cpp, which we do not cover in this evaluation.

\subsection{Results}

\begin{figure*}[!t]
    \centering
    \begin{minipage}{.25\textwidth}
        \centering
        \includegraphics[width=0.99\linewidth]{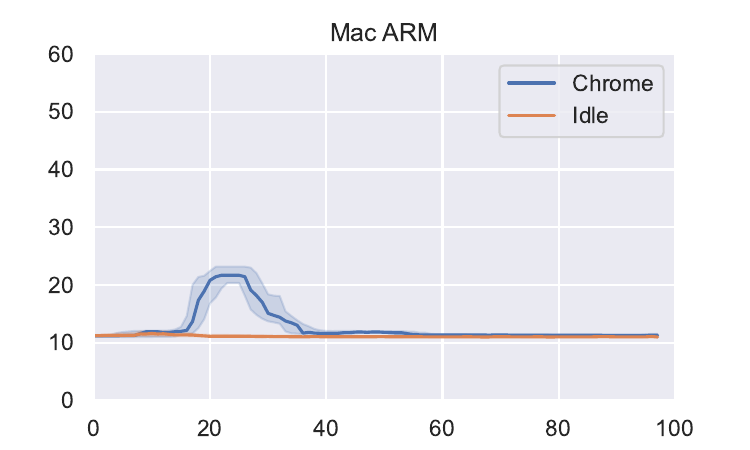}
        \label{fig:mac_arm}
    \end{minipage}%
    \begin{minipage}{.25\textwidth}
        \centering
        \includegraphics[width=0.99\linewidth]{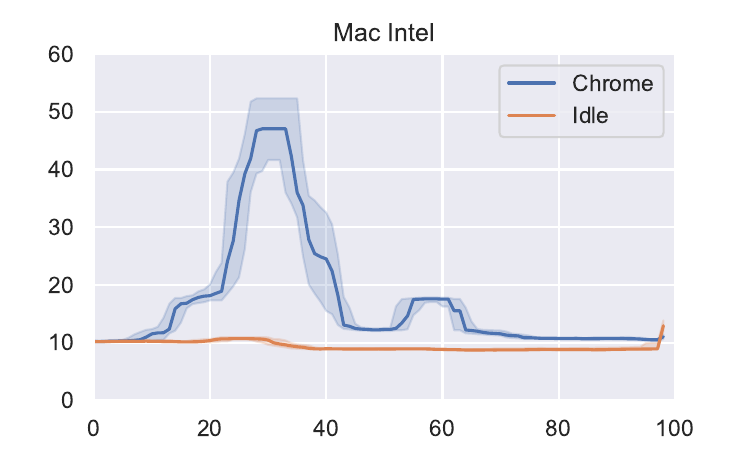}
        \label{fig:mac_intel}
    \end{minipage}%
    \begin{minipage}{.25\textwidth}
        \centering
        \includegraphics[width=0.99\linewidth]{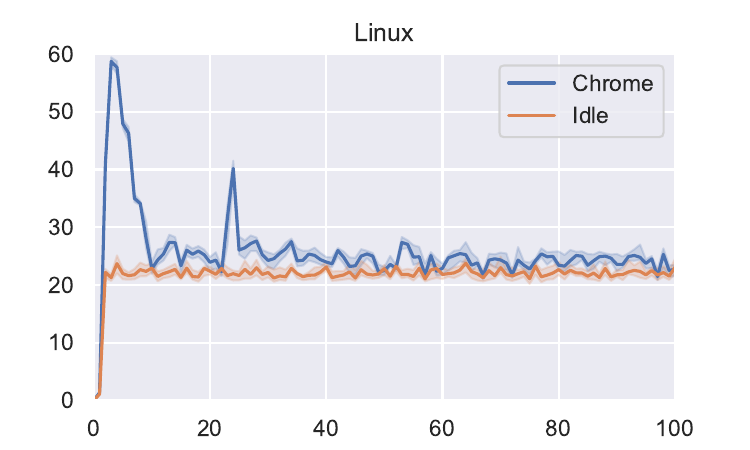}
        \label{fig:linux}
    \end{minipage}%
    \begin{minipage}{.25\textwidth}
        \centering
        \includegraphics[width=0.99\linewidth]{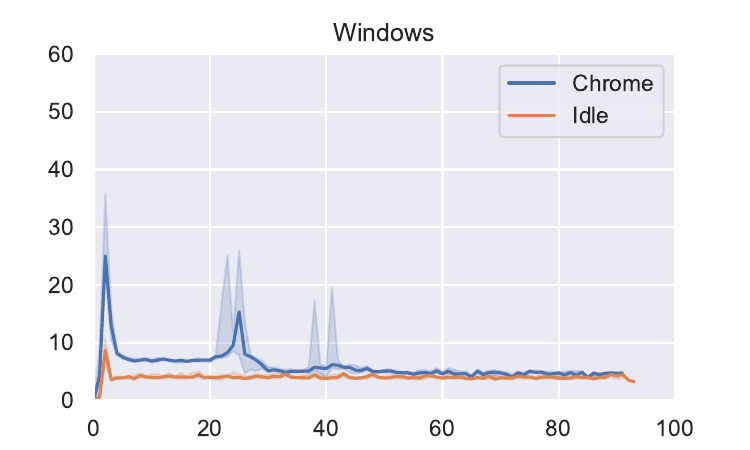}
        \label{fig:windows}
    \end{minipage}%
    \vspace{-1em}
    \caption{Evolution of the power consumption of Chrome and at IDLE on the four systems. The horizontal axis represents time (sample indexes) and the vertical axis represents measured power consumption in watts.}\label{fig:results}
    
\end{figure*}

In this section, we present the findings from our evaluation. \autoref{fig:results} provides a visual representation of energy consumption, via the power metric, for the Chrome browser and the system in an IDLE state across the four execution configurations (cf.~\autoref{tab:evaluation_harware}). The blue area illustrates the interquartile range for the 20 runs, while the line represents the mean power consumption. The vertical axis is measured in watts, and the horizontal axis represents the time evolution, indicated by the index of energy measurements, \ie samples, taken (one measurement every 100 milliseconds).

Significant differences are evident when comparing the four systems. Notably, the most prominent difference is the scale of power consumption between the systems. 
Power consumption on Mac ARM is notably lower than that on Mac Intel, the peak is around 20 watts for ARM while the peak is around 50 watts on Intel. 
We observe that the Linux system is the most consuming hardware configuration and the most powerful one too.
However, the overall power consumption is the lowest on Windows. 
This difference arises from the fact that on Windows and Linux, we measure CPU consumption, while on Mac, we measure the entire system's energy consumption. 
A notable observation is that the overall energy consumption of the entire Mac system is lower than the CPU consumption of the Linux machine.
Where the idle power consumption on Linux is around 20 watts while the peak on Mac ARM is 20 watts. Note however that we do not use equivalent hardware across different platforms -- we provide these observations to showcase the tool and the differences in the studied platforms are not generalizable.

Comparing the results between the IDLE state and Chrome execution, it is evident that the energy consumption peak occurs when Chrome is launched. Notably, the timing of this peak varies across the systems. On Linux, the peak occurs at the beginning, while on Mac, it takes place after approximately 20 samples, and on Windows, it is observed at the beginning and a second one around index 30. This discrepancy highlights the distinct behavior of each operating system.

\vspace{.2cm}{\centering\setlength{\fboxrule}{0.1pt}\fbox{\colorbox{mygray}{\parbox{0.95\columnwidth}{
During this evaluation, we show the ability of \tool to run on multiple platforms that have a diversity of software and hardware.
We demonstrate that interesting patterns can be identified by comparing the execution of Chrome on different platforms.
We also aimed to illustrate that \tool is easy to use and enables students, researchers, and practitioners to easily collect energy data from their software execution.
}}}}
\vspace{-1em}
\section{Threat to Validity}
Our research and the use of \tool for energy measurement and system metrics analysis are subject to several potential threats to validity.

Internal Bugs in \tool: An inherent concern is the possibility of internal bugs within \tool that may affect the accuracy of energy measurements and system metrics. To counter this threat, we have made \tool open-source, inviting developers and researchers to inspect and contribute to its codebase. We commit to ongoing maintenance and improvement to ensure its reliability. Furthermore, we intend to use \tool in the next edition of our course on Sustainable Software Engineering, where the tool will be used in a diverse set of platforms.

Impact of \tool Execution on Energy Measurement: The execution of \tool itself has the potential to impact energy measurements, particularly on certain platforms. \tool exhibits low and consistent energy consumption, thereby minimizing its influence on comparative analyses between two measurements.

Accuracy of Energy Measurements: Another potential threat relates to the accuracy of energy measurements. Ensuring precise measurements is challenging, and variations may occur in particular between systems. This is a limitation of any state-of-the-art software-based energy profiler. We argue that data collected from energy profilers yields results that are still comparable between executions under the same environment. This comparability allows \tool to effectively evaluate energy consumption trends within the measured system over time.

\vspace{-0.5em}
\section{Conclusion}

In this paper, we introduce \tool, the first energy measurement tool that is compatible with Linux, MacOS, and Windows, across a wide range of CPU architectures. \tool distinguishes itself by relying on low-level system calls to gather energy metrics, while maintaining low resource consumption. It also offers native support for measuring energy consumption on Nvidia GPUs, a feature often absent in other tools.

Our objective in providing \tool is to simplify the process of collecting energy measurements for students, researchers, and practitioners. The existing landscape for this task is currently complex and varies significantly across platforms, making it challenging to convey and establish a standardized approach to energy consumption measurement.

In our evaluation, we demonstrate \tool's capability to collect energy data across diverse operating systems and hardware setups. We aim to emphasize that this process can be relatively straightforward and consistent, irrespective of the platform.

While \tool lays a robust foundation for cross-platform energy measurement, there is ample room for further enhancement and expansion. In future work, we plan to:

\begin{itemize}[leftmargin=10pt]
    \item Explore additional metrics and parameters that can provide deeper insights into software energy consumption.
    \item Extend support for additional hardware configurations and provide scripts for in-depth analysis of the collected data.
    \item Incorporate hardware measurements for the entire system and synchronize them with software-based measurements.
\end{itemize}

\bibliography{references}
\end{document}